\documentclass[aps,prb,twocolumn,amsmath,superscriptaddress,groupedaddress]{revtex4}  
\usepackage{graphicx}  
\usepackage{dcolumn}   
\usepackage{bm}        
\usepackage{amssymb}   

\begin{document}

\title{Edge reconstruction of fractional quantum Hall liquids with spin degrees of freedom}

\author{Yuhui Zhang}
\affiliation{National High Magnetic Field Laboratory and
  Department of Physics, Florida State University, Tallahassee, FL
  32306, USA}

\author{Zi-Xiang Hu}
\affiliation{ Department of Physics, Chongqing University, Chongqing 400044, China}

\author{Kun Yang}
\affiliation{National High Magnetic Field Laboratory and
  Department of Physics, Florida State University, Tallahassee, FL
  32306, USA}

\date{\today}

\begin{abstract}

We study the interplay of confining potential, electron-electron interaction, and Zeeman splitting at the edges of fractional quantum Hall liquids, using numerical diagonalization of finite-size systems. The filling factors studied include 1/3, 5/2, 2/5, and 2/3. In the absence of Zeeman splitting and an edge, the first two have spin fully polarized ground states, while the latter two have singlet ground states. We find that with few exceptions, edge instabilities of these systems are triggered by softening of edge spin waves for Abelian fractional quantum Hall liquids ($1/3$, $2/5$ and $2/3$ liquids), and are triggered by softening of edge magnetoplasmon excitations for non-Abelian $5/2$ liquid at the smoother confinement side. Phase diagrams are obtained in the accessible parameter spaces.

\end{abstract}

\maketitle

\section{Introduction}

Quantum Hall edge states are not only responsible for the dissipationless quantized Hall transport, but also provide a unique window to study the highly non-trivial topological properties of the bulk. For example, it was predicted that for a large class of fractional quantum Hall (FQH) states, their edges exhibit {\em universal} low-energy properties that directly reflect the bulk topological order.\cite{Wenbook} On the other hand various complications, generally referred to as edge reconstruction, can happen at the edge due to interplay between confining potential (which tends to hold electrons together) and Coulomb repulsion between electrons (that tends to spread out electron charge).\cite{Glazman} It was shown in earlier numerical work\cite{Wan,JainNonlinear,murthy} that FQH edges are much more susceptible to reconstruction as compared to their integer counterparts. When edge reconstruction occurs gapless edge modes {\em not} required by bulk topological order appear, and they can ruin the predicted universality.\cite{KunEffectiveTheory}

Earlier work\cite{Wan,JainNonlinear} on FQH edge reconstruction has mostly ignored the electron spin degree of freedom (perhaps the only exception is Ref. \onlinecite{hu}). On the other hand it is known that spin can play a very active role at the edges of integer quantum liquids.\cite{Dempsey,Oaknin} Motivated by this we study the effects of confining potential, electron-electron interaction, and Zeeman splitting on the charge and spin structures at the edge of FQH liquids, by carrying out exact diagonalization calculations on small electron systems using disk geometry. The model and numerical methods are similar as the ones used in our earlier work on the spin structure of integer quantum Hall edges.\cite{Yuhui} More specifically, we study FQH liquids with filling factors $\nu=1/3$, $2/5$ and $2/3$, subject to positive background charge confining potential, briefly reviewed in Sec. II. We also study filling factor $\nu=5/2$, which is of very strong current interest. In this case we use a model similar to Refs. \onlinecite{wan06, wan08}, but also include spin degrees of freedom. Details will be discussed in Sec. II.

In our numerical studies, edge reconstruction is triggered by a level crossing between the ground state and an edge excited state (before reconstruction), or equivalently, softening of an edge mode. Depending on whether this (softened) excitation belongs to charge or spin edge mode, we can distinguish the reconstruction as charge or spin edge reconstructions. Therefore we need to know about all the edge excitations of a FQH liquid before studying how they lead to edge reconstructions. Within the framework of composite fermion (CF) theory\cite{Jain}, we can map some of the FQH filling factors to corresponding integer quantum Hall fillings of CFs, and use knowledge and intuition obtained from earlier extensive studies of these integer quantum Hall states' edges. For example spin-polarized $1/3$ state can be mapped onto filling factor $1$ of composite fermions. Therefore we expect that it has a non-chiral edge spin wave (ESW) mode and a forward-moving (chiral) edge magnetoplasmon (EMP) mode as the $\nu=1$ spin-polarized state.\cite{Yuhui} Spin-unpolarized $2/5$ and $2/3$ FQH states are mapped onto filling factor $2$ of composite fermions, with effective magnetic field parallel and antiparallel to the original external field respectively. As a result $2/5$ state has one forward-moving ESW mode and one forward-moving EMP mode as $\nu=2$ state.\cite{Jain2/3} The situation of the $2/3$ state is less trivial, and detailed study\cite{Jain2/3} shows that it has one backward-moving ESW mode and one forward-moving EMP mode. At $\nu = 5/2$, we find electrons in the half-filled 1st excited Landau level condense into the spin-polarized Moore-Read Pfaffian state in certain range of confining strength. Similar to spin-polarized $1/3$ state, the $5/2$ state has a non-chiral ESW mode and a forward-moving (chiral) EMP. In addition, it also has a forward-moving (chiral) edge Majorana fermion (EMF) mode.\cite{edgeBasis, wen5/2} The intuitions of some qualitative properties of edge reconstruction, like the relation between the directions of the edge reconstructing mode and confining strength, can be obtained from an electrostatic model described in our previous work.\cite{Yuhui}

Our most robust results are summarized as follows. Without Zeeman coupling, spin-polarized $1/3$ Laughlin-like state, spin-unpolarized $2/5$, $2/3$ Halperin-like states and spin-polarized Moore-Read Pfaffian state appear as ground states in certain regions of confining strength at corresponding filling factors. The non-chiral ESWs of spin-polarized $1/3$ state can be mapped onto $\Delta S = -1$ bosons on top of the spin-polarized $1/3$ state; the chiral ESWs of spin-unpolarized $2/5$ and $2/3$ can be mapped onto the pure spin excitations predicted by SU($2$) effective theory.\cite{Moore} In the spectra obtained by exact diagonalization, these ESW modes are low-lying and well separated from the other edge charge modes and bulk excitations. For each Abelian FQH liquid ($\nu = 1/3$, $2/5$ or $2/3$), edge reconstruction is triggered by softening of ESW mode if the corresponding ESW mode exists, indicating their importance. On the other hand, for the spin-polarized Moore-Read Pfaffian state, we find spin plays {\em no} role in its instabilities when confining potential strength is varied within our model. We also find that it is an EMP excitation that reconstructs the Pfaffian state at the smoother confinement side. The critical parameters of all the instabilities are identified in accessible finite size systems, and also estimated for the thermodynamic limit. The effects of Zeeman coupling are discussed whenever appropriate.

The rest of the paper is organized as follows. In Sec. II, we introduce the models and provide some numerical details of the exact diagonalization calculation. In Sec. III, we study the edge excitations and instabilities of the Abelian FQH liquids ($1/3$, $2/5$ and $2/3$ FQH liquids) in finite size systems. Sec. IV considers the instabilities of the Pfaffian state in finite size systems. Some conclusions and remarks are offered in Sec. V.

\section{MICROSCOPIC MODELS}
We consider interplay of two-body Coulomb interaction, one-body rotationally invariant confining potential and Zeeman coupling in our problem. For studying the FQH states with electrons confined to the lowest Landau level (LLL) (relevant for $1/3$, $2/5$ and $2/3$ states), the complete Hamiltonian in the symmetric gauge is
\begin{equation}
\begin{split}
H =& \frac{1}{2}\underset{m,n,l,\sigma,\sigma'}{\sum}V_{mn}^{l}c_{m+l,\sigma}^{\dagger}c_{n,\sigma'}^{\dagger}c_{n+l,\sigma'}c_{m,\sigma}+\underset{m,\sigma}{\sum}U_{m}^{cp}\hat{n}_{m,\sigma} \\
&+\frac{1}{2}g\mu_{B}B\underset{m}{\sum}(\hat{n}_{m,\uparrow}-\hat{n}_{m,\downarrow}) ,
\end{split}
\label{completeH}
\end{equation}
where $c_{m,\sigma}^{\dagger}$ is the electron creation operator for the LLL single-electron state with orbital angular momentum $m$ and spin $\sigma$, $\hat{n}_{m,\sigma}=c_{m,\sigma}^{\dagger}c_{m,\sigma} $ is the occupation number operator of the $m$th orbital with spin $\sigma$. $\mu_{B}$ is the Bohr magneton and $g$ is the electron spin $g$ factor. $V_{mn}^{l}$ are the corresponding matrix elements of Coulomb interaction for the symmetric gauge, and $U_{m}^{cp}$ are the matrix elements of the rotationally invariant confining potential. We assume a uniformly distributed neutralizing positive background charge layer at a distance $d$ above the two-dimensional electron gas (2DEG) to model the real 2DEG's confinement of a modulation-doped AlGaAs/GaAs heterostructure. Therefore 
\begin{equation}
U^{cp}_{m}=\frac{e\rho}{2\pi2^{m}m!}\int\int_{r_{2}<R}d^{2}r_{1}d^{2}r_{2}\frac{1}{\sqrt{d^{2}+r_{12}^{2}}}r_{1}^{2m}e^{-r_{1}^{2}/2},
\end{equation}
where $\rho$ is the charge density of the background, and $\epsilon$ is the dielectric constant. $d/l_{B}$ is the dimensionless ratio that tunes the relative strength between confining potential and electron-electron interaction. In addition we assume there is a sharp cleaved edge,\cite{sharpEdge} and model its effect by restricting the LLL orbitals to those with angular momentum from $m=0$ to $m_{max}=N/\nu-1$ (we make $m_{max} = N/\nu$ when study the $2/3$ state for a reason discussed later). Additional details of this model can be found in Ref. \onlinecite{Wan} and also our previous work for integer quantum Hall states.\cite{Yuhui} Total angular momentum $M$, total spin $S$ and its z-axis component $S_{z}$ are good quantum numbers, because the Hamiltonian (\ref{completeH}) has rotational symmetry and also commutes with ${\bf S}^2$ and $S_{z}$.

To study the $5/2$ state, we use the same treatment as in Refs. \onlinecite{wan06,wan08}. We explicitly keep the electronic states in the half-filled first Landau level (1LL) and neglect the spin up and down electrons in the LLL by assuming that they are inert. The amount of charge in the positive background charge disk is chosen to neutralize the electrons' charge in the 1LL. The disk encloses exactly $2N$ magnetic flux quanta, in which $N$ is the number of electrons in the 1LL. With this simplification, the Hamiltonian used to study $5/2$ state still has the form of (\ref{completeH}). But there are some differences: $m$ labels the $m$th orbital in the 1LL; $V_{mn}^{l}$, $U_{m}$ are the corresponding matrix elements of Coulomb interaction and confining potential for the electronic states of the 1LL.

In GaAs, Coulomb interaction dominates the Zeeman coupling energy in the magnetic fields of interest. We will treat the Zeeman coupling in a similar way as our previous study of integer quantum Hall states.\cite{Yuhui} At first the Zeeman term is ignored when studying the edge spin excitations. So for each energy level with quantum number $S$, it has degeneracy $2S+1$ with different values of $S_{z}$. When the Zeeman term is added back, the eigenstates will not change, but the original degenerate energy levels will be split corresponding to different $S_{z}$. We will also consider the effects of such splitting in the following.

\section{EDGE EXCITATIONS AND RECONSTRUCTIONS OF ABELIAN FQH liquids}

In this section, we study the edge excitations and reconstructions of spin-polarized $1/3$ Laughlin-like state, and spin-unpolarized $2/5$, $2/3$ Halperin-like states. All these three FQH states appear as ground states in certain regions of corresponding parameter spaces. We also find that in the absence of Zeeman splitting, in all cases edge reconstruction is triggered by softening of ESW mode if the corresponding ESW exists. More specifically if forward-moving ESW exists in a FQH liquid, edge reconstruction is triggered by it at the smoother confinement side; if backward-moving ESW exists, edge reconstruction is triggered by it at the stronger confinement side.

\subsection{Spin-polarized $1/3$ state}
\begin{figure}[h]
\includegraphics[width=9cm]{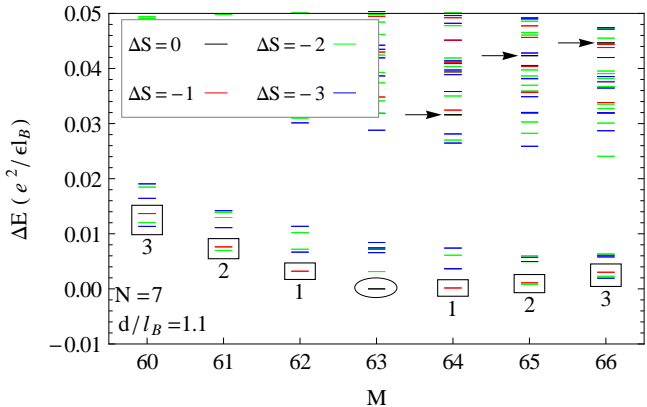}
\caption{(Color online) Low energy spectrum of $7$-electron system with $\nu = 1/3$ and $d/l_{B}=1.1$. The spin quantum numbers' differences compared to the spin-polarized state $\Delta S$ of the eigenstates are labeled by different colors as the annotation. The ground state is spin-polarized $1/3$ Laughlin-like state at $M=63$ (enclosed by a circle). Two branches of edge spin waves with opposite dispersions are enclosed by boxes. The number below each box is the number of states inside the box, and each state has degeneracy $2 S + 1$ based on its spin quantum number $S$. The spin-polarized edge magnetoplasmon excitations are labeled by arrows and merge into other bulk excitations because of high velocity.}\label{spectrum_v13_N7_pbg}
\end{figure}
\begin{figure}[h]
\includegraphics[width=8.5cm]{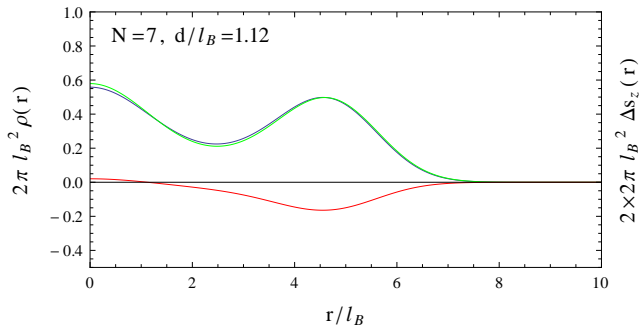}
\caption{(Color online) Charge density and spin density profiles of $1/3$ spin-polarized ground state and its edge spin wave with modulation angular momentum $\Delta M = 1$ at $d/l_{B}=1.12$ in the $7$-electron system. The {\em blue} and {\em green} lines are the normalized charge density functions $2\pi l_{B}^{2}\rho(r)$ of $\nu=1/3$ spin-polarized ground state and its $\Delta M = 1$ edge spin wave; the {\em red} line is z axis component of normalized spin density function $2 \times 2 \pi l_{B}^{2} \Delta s_{z}(r)$ of the $\Delta M = 1$ edge spin wave.}
\label{rdis_v13_N7}
\end{figure}

\begin{figure}[h]
\includegraphics[width=9cm]{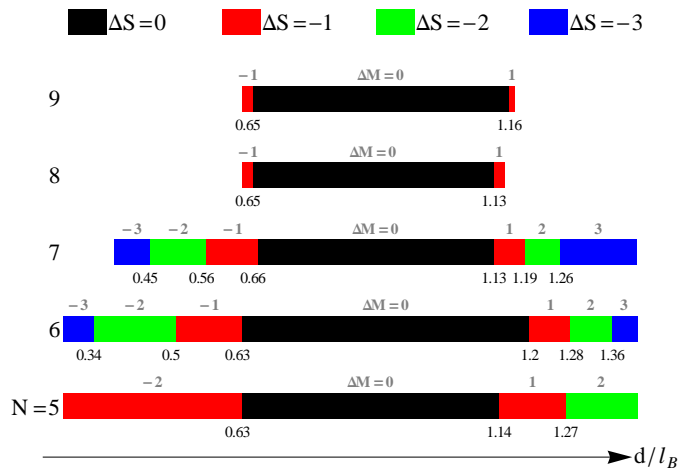}
\caption{(Color online) Phase diagrams of small electron systems with electron numbers $5-9$ at $\nu=1/3$. Spin-polarized states at 1/3 filling are distinguished by having the same angular momentum quantum number $M_{1/3}=3N(N-1)/2$ as the Laughlin state. $\Delta S=S-N/2$ ($\Delta M = M-M_{1/3}$) is the change of the total spin (angular momentum) compared to $\nu=1/3$ spin-polarized state. Due to limitation of growing size of Hilbert space, the phase diagrams of $8$- and $9$-electron systems are obtained by assuming that the state destabilizing $1/3$ state has $\Delta S=-1$. Different $\Delta S$ of ground states are labeled by different colors as in the annotation.}\label{phase_diagram_v13_pbg}
\end{figure}

In our numerical results, spin-polarized $1/3$ Laughlin-state is distinguished by having the same quantum numbers $M=M_{1/3} \equiv 3N(N-1)/2$, $S=N/2$ as the $1/3$ Laughlin state, and it appears as ground state in a region of parameter space. For brevity, we abbreviate this $1/3$ Laughlin-like state to $1/3$ state in the following. Similar to the spin-polarized $\nu = 1$ state,\cite{Yuhui} spin-polarized $1/3$ state has two branches of edge excitations. One of the two branches is the (non-chiral) ESW mode. In the spectrum, these ESWs are well separated from the other excitations. As shown in Fig. \ref{spectrum_v13_N7_pbg}, the ESW excitations enclosed by boxes obey boson counting in each subspace $M$, and the other low-lying states are combinations of the ESWs. The spin configuration in each subspace $M$ tells us that each ESW of $1/3$ state would change the system's total spin by $-1$. Through checking the charge and spin density profiles like those in Fig. \ref{rdis_v13_N7}, we verified that these ESWs do have spin textures localized at the edge. Besides the ESW mode, the well studied EMP mode merges into the bulk excitations in the spectra of small systems because of its high velocity.

With changing confining potential, the softening of ESWs reconstructs the edge of spin-polarized $1/3$ state at both smoother and stronger confinement sides as shown in Fig. \ref{phase_diagram_v13_pbg}.
Since each single $1/3$ state's ESW changes the system's total spin by $-1$, and the initial reconstructing state with $\Delta S = -1$ is a single ESW with modulation angular momentum $\Delta M = M - M_{1/3} = 1$ at the smoother confinement side, or $\Delta M = -1$ at the stronger confinement side. From the correlation of $S$ and $M$ of all the states appearing in the phase diagram (Fig. \ref{phase_diagram_v13_pbg}), we can deduce that with further smoothing (strengthening) confinement, the ESWs with $\Delta M = \pm 1$ will be generated one by one to reconstruct the $1/3$ state.
(There is one exception in $5$-electron system at the stronger confinement side because of the finite size effect of this very small system.) We also find with increasing particle number from $5$ to $9$, the two critical parameters $d_{c_{1}}/l_{B}$ and $d_{c_{2}}/l_{B}$ of edge reconstructions have small fluctuations $\sim0.01 $ as shown in Fig. \ref{phase_diagram_v13_pbg}. These enable us to predict that in thermodynamic limit $d_{c_{1}}/l_{B}$ is between $0.6$ and $0.7$ while $d_{c_{2}}/l_{B}$ is between $1.1$ and $1.2$. The $1/3$ state is stable when $d_{c_{1}}/l_{B}<d/l_{B}<d_{c_{2}}/l_{B}$.

The low-lying excitations' pattern of spin-polarized $1/3$ state is identical to that of spin-polarized $\nu = 1$ state for small systems in our previous numerical study.\cite{Yuhui} This is not a surprise, because in composite fermion theory, $\nu = 1/3$ state of electrons can be viewed as $\nu=1$ state of composite fermions under a (reduced) effective magnetic field. One difference about the reconstructions is that the backward-moving ESW can destabilize the spin-polarized $1/3$ state at the stronger confinement side, while the spin-polarized $\nu=1$ state is stable as parameter $d/l_{B}$ approaches zero. This difference originates from a quantitative difference, namely the electrons in spin-polarized $\nu=1$ state has stronger exchange effect, resulting in stronger stability of the polarized state.

\begin{table}[h]
\begin{tabular}{|c||c|c|c|c|c|}
\hline \hline
$N$  & $\tilde{g_{c}}$ & $B_{c}$($T$) & $d_{c}/l_{B}$\\ \hline
5 & 0.0034  & 0.34 & 1.44 \\ \hline
6 & 0.0038  & 0.43 & 1.5  \\ \hline
7 & 0.0025  & 0.19 & 1.35 \\ \hline
8 & 0.0022  & 0.14 & 1.36 \\ \hline
9 & 0.0029  & 0.25 & 1.44 \\
\hline \hline
\end{tabular}
\caption{Critical values of {\em charge} edge reconstruction of spin-polarized $1/3$ state in finite size systems. $\tilde{g_{c}}=g_{c} \mu_{B} B/(e^{2}/\epsilon l_{B}) $ is the normalized critical $g$ factor in which charged edge magnetoplasmon instead of neutral edge spin wave becomes the initial instability of spin-polarized $\nu=1/3
$ state at critical parameter $d_{c}/l_{B}$; $B_{c}$ is the magnetic field corresponding to $\tilde{g_{c}}$ in a AlGaAs/GaAs heterostructure with dielectric constant $\epsilon=12.8$.}
\label{1/3table}
\end{table}

Finite Zeeman term increases the energies of ESWs compared to the spin-polarized $1/3$ FQH state. The dimensionless parameter $\tilde{g}=g \mu_{B} B/(e^{2}/\epsilon l_{B}) $ is the ratio of the Zeeman energy to the typical Coulomb energy. For large enough $\tilde{g}=\tilde{g_{c}}$ the spin-polarized state will be destabilized by its spin-polarized excitations at the smoother confinement side, and softening of EPM will replace ESW to become the initial instability of spin-polarized state. The critical values of this transition are shown in Table \ref{1/3table}. The normalized critical $g$ factor has the same order $10^{-3}$ as that of $\nu=1$ state.\cite{Yuhui} In a AlGaAs/GaAs heterostructure with dielectric constant $\epsilon=12.8$, such critical $g$ factor corresponds to the magnetic field strength $\sim 0.1 T$. We thus conclude that unless the $g$ factor is tuned to be very close to zero, edge instability at the smoother confinement side will be triggered by charge reconstruction in typical samples.

\subsection{Spin-unpolarized $2/5$ state}
\begin{figure}[h]
\includegraphics[width=9cm]{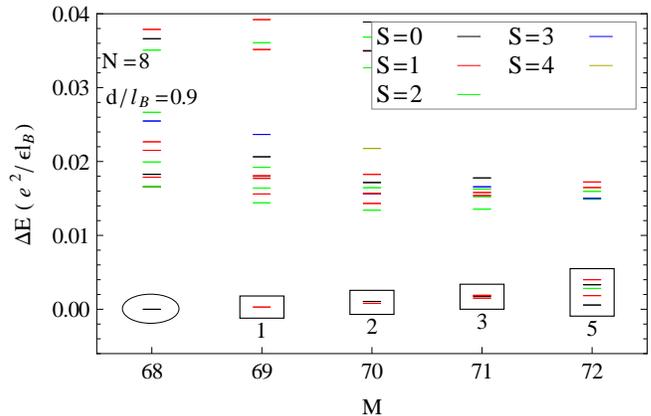}
\caption{(Color online) Low energy spectrum of an $8$-electron system at $\nu=2/5$ and $d/l_{B}=0.9$. The different total spin quantum numbers $S$ of the eigenstates are labeled by different colors as in the annotation. The ground state is spin-unpolarized $2/5$ Halperin-like state at $M=68$ (enclosed by a circle). Low energy spin excitations are enclosed by boxes. The number below each box is the number of states inside the box, and each state has degeneracy $2 S + 1$ based on its spin quantum number $S$.}\label{spectrum_v25_pbg}
\end{figure}

In our numerical results, spin-unpolarized $2/5$ Halperin-like state is distinguished by having the same quantum number $M_{2/5}=N(5N-6)/4$ and $S = 0$ as the $2/5$ Halperin state\cite{Halperin}
\begin{equation}
\begin{split}
\Psi^{(H)}_{2/5} = & \underset{i<j}{\prod}(z_{\uparrow i}-z_{\uparrow j})^{3}\underset{i<j}{\prod}(z_{\downarrow i}-z_{\downarrow j})^{3}\underset{i,j}{\prod}(z_{\uparrow i}-z_{\downarrow j})^{2} \\
&\times e^{-\frac{1}{4l_{B}^{2}}(\sum_{i}|z_{\uparrow i}|^{2}+\sum_{j}|z_{\downarrow j}|^{2})} , \label{25Halperin}
\end{split}
\end{equation}
where $z_{\sigma i}$ is the complex coordinate of the $i$th electron with spin $\sigma$ ($\sigma= \uparrow$ or $\downarrow$). This $2/5$ Halperin-like state appears as ground state from $d=0$ to critical parameter $d_{c}$ in our exact diagonalization results. It also has a large overlap with the $2/5$ Halperin state (Eq. \ref{25Halperin}) ($0.92$ in $6$-electron system and $0.84$ in $8$-electron system at $d/l_{B} = 0.9$). For brevity, we abbreviate this $2/5$ Halperin-like state to $2/5$ state in the following. In the spectrum Fig. \ref{spectrum_v25_pbg}, some low energy excitations in subspaces $\Delta M=M-M_{2/5} > 0$ are well separated from the other excitations. By comparing these excitations' spin quantum numbers in each subspace $\Delta M$ with the ones predicted by SU($2$) effective theory,\cite{Moore} we verified that these excitations are pure spin excitations and constitute the (forward-moving) chiral ESW branch. The other polarized EMP excitations have high velocity and merges into bulk excitations in small systems. They can be distinguished by calculating the overlaps to corresponding composite fermion wave functions.\cite{Jain2/3} We will not distinguish the EMP mode of $2/5$ state in this work because in the $M$ subspaces close to $2/5$ state, ESWs always have much lower energies and thus are the ones reconstructing the edge of $2/5$ state.

Through exact diagonalization, we find that the critical parameters $d_{c}/l_{B}$ at which the spin-unpolarized $2/5$ state is reconstructed by softening of ESW is $0.92$ for both $6$- and $8$-electron systems. $d_{c}/l_{B}$ for larger systems are not accessible due to numerical difficulty. Previous numerical works show that the critical points of FQH liquids' edge reconstructions converge to thermodynamic limit very fast.\cite{Wan,JainNonlinear} Based on this experience and consistency of results from $6$- and $8$-electron systems, we expect that in thermodynamic limit, at critical parameter $d_{c}/l_{B} \approx 0.9$, spin-unpolarized $2/5$ state is reconstructed by softening of its ESW. Different from the case in $1/3$ state, the ESW excitation initially reconstructing $2/5$ state is not the one with minimum angular momentum $\Delta M = 1$, but the one in subspace $\Delta M = 3$ ($\Delta S = 0$) for $6$-electron system and $\Delta M = 4$ ($\Delta S = 0$) for $8$-electron system. Given more orbital numbers, $\Delta M$ of the initially reconstructing ESW state will not decrease, because the energy of the state with larger $\Delta M$ decreases more as the orbital number increases.

The low-lying excitations' pattern of spin-unpolarized $2/5$ state is identical to the one of $\nu=2$ spin-unpolarized state for small systems in our previous numerical study.\cite{Yuhui} This can be easily understood by composite fermion theory, because $\nu=2/5$ state of electrons can be mapped onto $\nu=2$ state of composite fermions under a (reduced) effective magnetic field. In both cases, without Zeeman coupling, softening of ESW mode triggers the initial edge reconstruction of QH liquid.

In thermodynamic limit since the values of momentum becomes continuous and the ESW mode is gapless, the ESW mode (with finite spin quantum number) will destabilize the singlet $2/5$ state with any finite Zeeman coupling. We note in passing that edge reconstruction of the spin fully polarized 2/5 state has been studied before.\cite{murthy}

\subsection{Spin-unpolarized 2/3 state}
\begin{figure}[h]
\includegraphics[width=9.2cm]{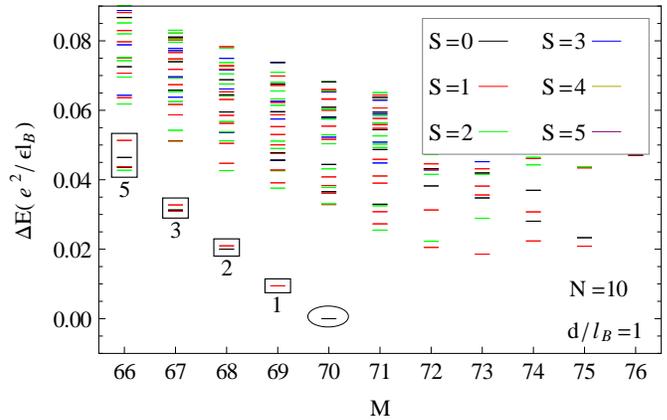}
\caption{(Color online) Low energy spectrum for a $10$-electron system at $\nu=2/3$ and $d/l_{B}=1.0$. The $10$ electrons are confined in $16$ orbitals. The different total spin quantum numbers $S$ of the eigenstates are labeled by different colors as in the annotation. The ground state is spin-unpolarized $\nu=2/3$ Halperin-like state at $M=70$ (enclosed by a circle). Low energy spin excitations are enclosed by boxes. The spin-polarized edge magnetoplasmon excitations merge into the bulk excitations because of high velocity. With stronger confinement, the ESW excitation will reconstruct the spin-unpolarized $2/3$ Halperin-like state. With smoother confinement, the lowest state (quasihole) at subspace $M = 75$, $S = 1$ will destabilize the spin-unpolarized $2/3$ Halperin-like state.}\label{spectrum_v23_pbg}
\end{figure}

In our numerical results, spin-unpolarized $2/3$ Halperin-like state is distinguished by having the same quantum number $M_{2/3}=N(3N-2)/4$ as the $2/3$ Halperin state\cite{Halperin}
\begin{equation}
\begin{split}
\Psi_{2/3}^{(H)} = & \underset{i<j}{\prod}(z_{\uparrow i}-z_{\uparrow j})\underset{i<j}{\prod}(z_{\downarrow i}-z_{\downarrow j})\underset{i,j}{\prod}(z_{\uparrow i}-z_{\downarrow j})^{2}\\
& \times e^{-\frac{1}{4l_{B}^{2}}(\sum_{i}|z_{\uparrow i}|^{2}+\sum_{j}|z_{\downarrow j}|^{2})} ,
\end{split}
\label{H23wf}
\end{equation}
where $z_{\sigma i}$ is the complex coordinate of the $i$th electron with spin $\sigma$ ($\sigma = \uparrow$ or $\downarrow$).  The $2/3$ Halperin-like state appears as ground state in a region of parameter space. Unlike $2/5$ Halperin-like state, this $2/3$ Halperin state obtained by the Coulomb Hamiltonian (\ref{completeH}) has a very small overlap with the $2/3$ Halperin state (\ref{H23wf}), which is $0.1702$, $0.0154$, $0.0004$ for $6$-, $8$-, $10$-electron system respectively at $d/l_{B} = 1.0$. The main reason of this small overlap is that the $2/3$ Halperin state does not have a good spin quantum number $S$. \cite{2/3noteI} Compared with the $2/3$ Halperin state, spin-unpolarized $2/3$ composite fermion state (with the same angular momentum $M$ and topological property as Halperin state) has a good quantum $S$ and a much larger overlap with the Coulomb state.\cite{Jain2/3}

In effective theory\cite{WenEffectiveTheory}, the $K$ matrix of $2/3$ state is given by
\begin{equation}
K_{2/3}=\left(\begin{array}{cc}
1 & 2\\
2 & 1
\end{array}\right) .
\end{equation}
Its one positive and one negative eigenvalue imply one forward-moving and one backward-moving edge mode. Although not intuitive, detailed study of composite fermion theory can also derive one forward-moving EMP mode and one backward-moving ESW mode.\cite{Jain2/3} In spectrum Fig. \ref{spectrum_v23_pbg}, some low energy excitations in subspaces $\Delta M=M-M_{2/3} < 0$ are well separated from the other excitations. By checking these excitations' spin quantum numbers in each subspace $\Delta M$ with the ones predicted by SU($2$) effective theory,\cite{Moore} we verified that these excitations are pure spin excitations and constitute the (backward-moving) chiral ESW branch.\cite{2/3noteII} The forward-moving chiral EMP excitations with $\Delta M > 0$ have high velocity and merges into bulk excitations in small systems.

\begin{table}[h]
\begin{tabular}{|c||c|c|c|}
\hline \hline
$N$ & $d_{c1}/l_{B}$ & $d_{c2}/l_{B}$ \\ \hline
$6$ & $0.7$ & $1.26$ \\ \hline
$8$ & $0.69$ & $1.23$ \\ \hline
$10$& $0.68$ & $1.21$ \\
\hline \hline
\end{tabular}
\caption{Critical parameters of spin-unpolarized $2/3$ Halperin-like state's edge reconstructions in finite size systems. $d_{c1}/l_{B}$ is the critical parameter in which (backward-moving) edge spin wave (ESW) reconstructs the spin-unpolarized $2/3$ Halperin-like state; $d_{c2}/l_{B}$ is the critical parameter in which quasihole excitation destabilize the spin-unpolarized $2/3$ Halperin-like state.}
\label{2/3dcTable}
\end{table}

In our numerical calculation, $3N/2 + 1$ orbitals are given to each system so that the $2/3$ spin-unpolarized state's initial destabilizing states will not change along with further increasing orbital number. With {\em stronger} confinement, the {\em backward-moving} ESW will soften and destabilize the spin-unpolarized $2/3$ state at a critical parameter $d_{c1}/l_{B}$ (shown in Table \ref{2/3dcTable}). The ESW excitation initially reconstructing $2/3$ state is the one with modulation angular momentum $\Delta M = -1$.
With smoother confinement, the $2/3$ state is destabilized by a state with $S = 1$ and $\Delta M = N/2$ in the finite size systems at a critical parameter $d_{c2}$ (shown in Table \ref{2/3dcTable}). These quantum numbers are the same as a charge $-e/3$ quasihole with spin up or down.\cite{quasi-holeExplain} On the next paragraph we will verify that the instability at the smoother confinement side is a quasihole located at the center of the $2/3$ liquid.

\begin{figure}[h]
\includegraphics[width=8.3cm]{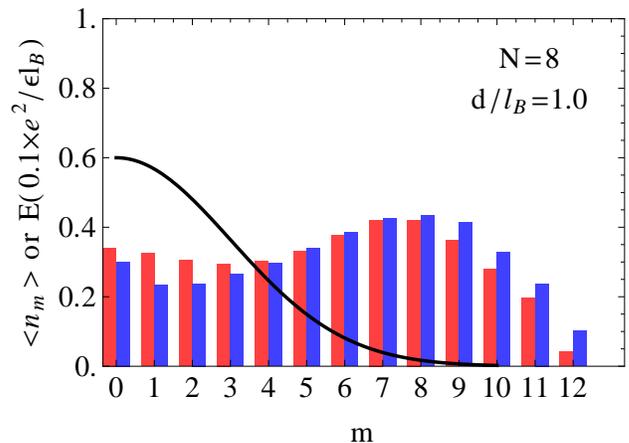}
\caption{(Color online) Orbital occupation number $\left\langle n_{m}\right\rangle $ of a single spin component for $2/3$ spin-unpolarized ground state (red bar) at $d/l_{B} = 1.0$ in $8$-electron system, and its quasihole instability with angular momentum $M$, spin $S = 1$, $z$-component of spin $S_{z} = 0$ (blue bar) generated by an extra Gaussian impurity potential with width $s = 3$ and amplitude $W_{g} = 0.06(e^{2}/\epsilon l_{B})$ (black curve). Although the Gaussian impurity potential is a discrete function of orbital number $m$, we plot it as a continuous function in this figure.}\label{ldis_v23}
\end{figure}

Besides smoothing the confinement, the same instability (at subspace $\Delta M = N /2$, $S = 1$) can be also excited by adding a Gaussian impurity potential $H_{W}$ on $2/3$ state:\cite{GaussianImpurity}
\begin{equation}
H_{W} = \sum_{m}U_{m}^{g}c^{\dagger}_{m}c_{m} ,
\end{equation}
and
\begin{equation}
U_{m}^{g} = W_{g} \exp(-m^2/2s^2),
\end{equation}
where $W_{g}$ with the unit $e^2/\epsilon l_{B}$ is the amplitude, and the dimensionless $s$ is the width of the Gaussian impurity. Take the $8$-electron system under a confining potential with $d/l_{B} = 1.0$ in which spin-unpolarized $2/3$ state is ground state as an example (Fig. \ref{ldis_v23}). If we add the Gaussian impurity potential $H_{W}$ with width $s = 3$ and increase the amplitude $W_{g}$ from $0$ to $0.06$($e^2 / \epsilon l_{B}$), the $2/3$ state will be destabilized by a quasihole at the center of the electron droplet (Fig. \ref{ldis_v23}). This quasihole is in the same subspace $\Delta M = 4$, $S = 1$, and also has a large overlap ($0.9923$ for their $S_{z} = 0$ wave functions) with the state without Gaussian impurity potential which is $2/3$ state's destabilizing state we observed above. In $6$- and $10$-electron systems, the destabilizing state can be also excited by adding a Gaussian impurity potential with slightly different width $s$ and amplitude $W_{g}$. For this reason, we conclude that in finite size systems the destabilizing state of spin-unpolarized $2/3$ state at the smoother confinement side is actually a charge $-e/3$ quasihole with a certain spin located at the center of the electron droplet. In thermodynamic limit, edge excitations' energies are lowered and will replace it to destabilize the $2/3$ spin-unpolarized state. Therefore the critical parameter $d_{c2}$ in which the $2/3$ state is destabilized obtained in finite size system (as shown in Table. \ref{2/3dcTable}) is not reliable when we consider the systems in thermodynamic limit.

\section{EDGE EXCITATIONS AND RECONSTRUCTIONS OF $5/2$ NON-ABELIAN FQH LIQUIDS}

\begin{figure}[h]
\includegraphics[width=9cm]{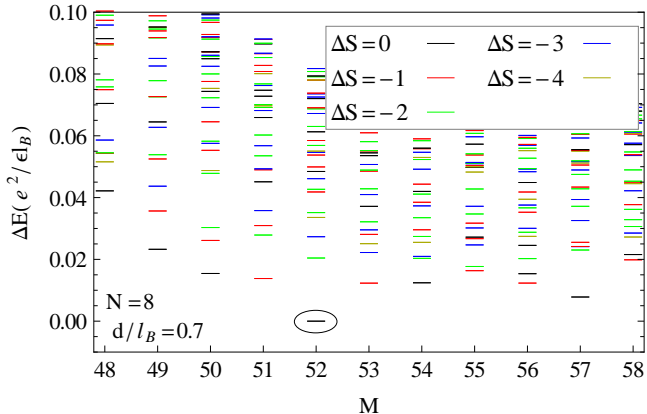}
\caption{(Color online) Low energy spectrum of $\nu=5/2$ for $8$-electron in $16$ orbitals with Coulomb two-body interaction and confining potential at $d/l_{B}=0.7$. $\Delta S = S - N/2$ is the spin quantum number's difference compared to the spin-polarized state. The ground state is spin-polarized Pfaffian state at $M=52$ (enclosed by a circle). It is hard to distinguish the edge excitations because they mix with the bulk excitations in finite size system. With smoother confinement, the state destabilizing the Pfaffian state is the low lying polarized state in subspace $M=57$.}\label{spectrum_v52_pbg}
\end{figure}

\begin{table}[h]
\begin{tabular}{|c||c|c||c|c|}
\hline \hline
$N$ & $d_{c1}/l_{B}$ & $\Delta M_{1}$ & $d_{c2}/l_{B}$ & $\Delta M_{2}$ \\ \hline
$6$  & NA      & NA   & $0.8$   & $4$  \\ \hline
$8$  & $0.01$  & $-8$ & $0.78$  & $5$  \\ \hline
$10$ & $0.55$  & $-4$ & $0.58$  & $5$  \\ \hline
$12$ & $0.47$   & $-5$ & $0.7$   & $6$  \\
\hline \hline
\end{tabular}
\caption{Critical parameters of the Pfaffian state's destabilizations in finite size systems with orbital numbers $2N$. $d_{1(2)c}/l_{B}$ is the critical parameter in which the Pfaffian state is destabilized at the stronger (smoother) confinement side. $\Delta M_{1(2)} \equiv M_{1(2)} - M_{P}$ is the angular momentum number's change compared to Pfaffian state's angular momentum $M_{P}=N(2N-3)/2$ for the states which initially destabilize the Pfaffian state. For $6$-electron system, Pfaffian state is stable as $d/l_{B}$ approaches zero, so $d_{1}/l_{B}$ is not applicable (NA) in this system. The spin of the destabilizing states is $\Delta S \equiv S- N/2 = 0$ (polarized). For numerical difficulty, only $-2\leq \Delta S \leq 0$ states are calculated in $10$-electron system and only $-1 \leq \Delta S \leq 0$ states are calculated in $12$-electron system.}
\label{5/2Table}
\end{table}

In our numerical result, spin-polarized Moore-Read Pfaffian state is distinguished by having the same angular momentum quantum number $M_{P}=N(2N-3)/2$ as the Pfaffian state proposed by Moore and Read for a half-filled LLL,\cite{MR}
\begin{equation}
\Psi_{Pf} = Pf(\frac{1}{z_{i}-z_{j}})\underset{i<j}{\prod}(z_{i}-z_{j})^{2}\exp(-\underset{i}{\sum}\frac{|z_{i}|^{2}}{4l_{B}^{2}}) ,
\end{equation}
where the Pfaffian is defined by
\begin{equation}
Pf(\frac{1}{z_{i}-z_{j}})=\frac{1}{2^{N/2}(N/2)!}\mathcal{A}(\frac{1}{z_{1}-z_{2}}\frac{1}{z_{3}-z_{4}}...) ,
\end{equation}
in which $\mathcal{A}$ is the anti-symmetrization operator. Since in the Pfaffian state electrons form p-wave pairs, we only study the finite size systems with even numbers of electrons. The Pfaffian state appears as ground state in a region of parameter space. In the low energy spectra of $\nu=5/2$ small systems, finite size effect is so serious that all the edge excitations: non-chiral ESW, forward-moving (chiral) EMP and edge Majorana fermion (EMF) excitation mix with the bulk ones as shown in Fig. \ref{spectrum_v52_pbg}. For this reason, it is hard to distinguish the edge excitations of the Pfaffian state.

With stronger confinement, the Pfaffian state is destabilized by a spin fully polarized excitation at critical parameter $d_{c1}/l_{B}$. Both the critical parameter $d_{c1}/l_{B}$ and $\Delta M$ of the destabilizaing state change a lot as the particle number increases from $6$ to $12$ as shown in Table \ref{5/2Table}. Since all the spin-polarized edge excitations of the Pfaffian state are chiral with $\Delta M > 0$, the state destabilizing the Pfaffian state at the stronger confinement side in Table \ref{5/2Table} (with $\Delta M < 0$) is a bulk excitation. As shown in Ref. \onlinecite{wan08}, in the filling factor $\nu = 5/2$ some other bulk states compete with the Pfaffian state to become the ground state with changing confining potential. This competition is very sensitive to the particle number in finite size systems, which is the reason why the critical parameter $d_{c1}/l_{B}$ and $\Delta M$ of the destabilizaing state change a lot with different particle number $N$.

With smoother confinement, the Pfaffian state is destabilized by a polarized excitation at critical parameter $d_{c2}/l_{B}$ as shown in Table \ref{5/2Table}.
To figure out whether this destabilizing state is edge or bulk excitation, instead of pure Coulomb Hamiltonian, we consider the following mixed Hamiltonian\cite{wan08}
\begin{equation}
H_{mix} = \lambda H_{3B} + (1 - \lambda) H_{c},
\end{equation}
in which the parameter $\lambda$ ($0 \leqslant \lambda \leqslant 1$) interpolates smoothly between the limiting cases of a pure three-body Hamiltonian $H_{3B}$ ($\lambda = 1$) and a pure Coulomb Hamiltonian $H_{c}$ ($\lambda = 0$) (including electron-elctron interaction and confining potential). Pfaffian state is the exact zero energy eigenstate of the $H_{3B}$ with the minimum angular momentum, and its edge excitations also have zero energy with a gap to other bulk excitations. As shown in Refs. \onlinecite{3BInt} and \onlinecite{wan08}, the three-body Hamiltonian can be written in terms of projection operators:
\begin{equation}
H_{3B}=\underset{\mathcal{M}}{\sum}\underset{i<j<k}{\sum}\left|\psi_{\mathcal{M}}(i,j,k)\right\rangle \left\langle \psi_{\mathcal{M}}(i,j,k)\right| ,
\label{3BHamiltonian}
\end{equation}
in which $\mathcal{M}$ is the total angular momentum of a $3$-electron cluster.
For fermions, the normalized $3$-body wave function is
\begin{equation}
\psi_{\mathcal{M}}(z_{1},z_{2},z_{3})=B_{\mathcal{M}}(z_{1}+z_{2}+z_{3})^{\mathcal{M}-3}J(z_{1},z_{2},z_{3}) ,
\end{equation}
where $J(z_{1},z_{2},z_{3})=(z_{1}-z_{2})(z_{1}-z_{3})(z_{2}-z_{3}) $ and the normalization factor is
\begin{equation}
B_{\mathcal{M}}=\frac{1}{(2\pi)^{3/2}}\sqrt{\frac{3^{\mathcal{M}-4}}{2^{\mathcal{M}+2}(\mathcal{M}-3)!}} .
\end{equation}
In occupation space, the three-body interaction $H_{3B}$ has a rather simple form\cite{wan08}
\begin{equation}
H_{3B}=\underset{m_{1}>m_{2}>m_{3}}{\sum}\underset{m_{4}<m_{5}<m_{6}}{\sum}U(\{m_{i}\})c_{m_{1}}^{\dagger}c_{m_{2}}^{\dagger}c_{m_{3}}^{\dagger}c_{m4}c_{m5}c_{m6} ,
\end{equation}
and
\begin{equation}
U(\{m_{i}\})=V(m_{1},m_{2},m_{3})V(m_{4},m_{5},m_{6}) .
\end{equation}
With $\mathcal{M}=m_{1}+m_{2}+m_{3}$, the antisymmetric function
\begin{equation}
V(m_{1},m_{2},m_{3})=\sqrt{\frac{(\mathcal{M}-1)!}{2\times3^{\mathcal{M}}m_{1}!m_{2}!m_{3}!}}\mathcal{A} \{m_{2}m_{1}(m_{1}-1)\} ,
\end{equation}
and $\mathcal{A} $ is the antisymmetrizer in $m_{1}$, $m_{2}$ and $m_{3}$.
Explicit construction of the basis of Pfaffian state's edge excitations\cite{edgeBasis} shows that to obtain the  ``right" degeneracy of edge excitations in a subspace $\Delta M$ for the exact diagonalization calculation, a certain number of orbitals is needed. With pure three-body Hamiltonian $H_{3B}$ and $2N+4$ (instead of $2N$) orbitals, the degeneracies of zero energy levels (number of edge excitations) in subspaces $1 \leq \Delta M \leq 6$ will not be reduced by the finite orbital number. Even so, the number of edge excitations in some subspace $\Delta M$ are still reduced by the small particle number, because the generations of Majorana fermion require the destruction of electron pairs. The degeneracies of zero energy levels (number of edge excitations) in subspaces $1 \leq \Delta M \leq 6$ for small systems are shown in Table \ref{5/2Table1}.
\begin{table}[h]
\begin{tabular}{|c||c|c|c|c|c|c|}
\hline \hline
$\Delta M$     &   $1$   &   $2$   &   $3$   &   $4$  &   $5$  &  $6$   \\ \hline
$N = 6$        &   $1$   &   $3$   &   $5$   &   $9$  &   $13$ &  $21$  \\ \hline
$N = 8$        &   $1$   &   $3$   &   $5$   &   $10$ &   $15$ &  $25$  \\ \hline
$N = 10$       &   $1$   &   $3$   &   $5$   &   $10$ &   $16$ &  $27$  \\ \hline
$N = 12$       &   $1$   &   $3$   &   $5$   &   $10$ &   $16$ &  $28$  \\
\hline \hline
\end{tabular}
\caption{The degeneracies of Pfaffian state's edge excitations in subspaces $1 \leq \Delta M \leq 6$ for finite size systems with $2 N + 4$ orbitals. $N$ is the particle number; $\Delta M \equiv M - M_{P}$ is the angular momentum number's change compared to Pfaffian state's angular momentum $M_{P}=N(2N-3)/2$ for the edge excitations. The orbital numbers are chosen as $2 N + 4$, so that the finite orbital number will not revise edge excitations' degeneracies in subspaces $1 \leq \Delta M \leq 6$ compared with the ones in infinite orbital system. The different degeneracies in some subspaces $\Delta M$ come from the effect of small particle number. The degeneracies of $N=12$ system in subspaces $1 \leq \Delta M \leq 6$ are the same as the ones in thermodynamic limit (also with large enough orbital number). }
\label{5/2Table1}
\end{table}
\begin{figure}[h]
\includegraphics[width=9cm]{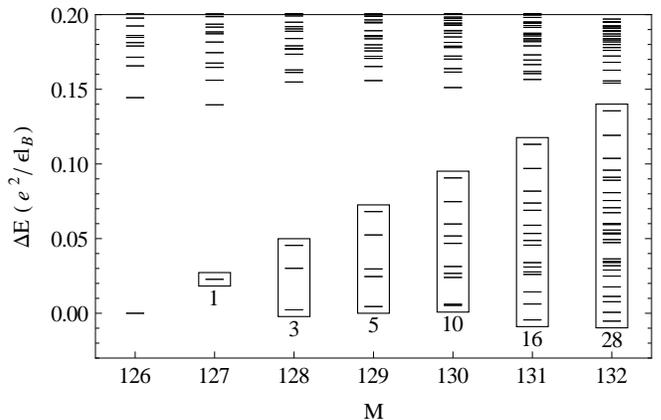}
\caption{Low energy spectrum of $\nu = 5/2$ for polarized $12$-electron in $28$ orbitals system with mixed Hamiltonian ($d/l_{B} = 0.7$, $\lambda = 0.5$). Unlike the pure Coulomb interaction case, edge excitations are well separated from the bulk ones (in this case some edge excitations in $\Delta M > 0$ already destabilize the Pfaffian state). They are enclosed by boxes. The number below each box is the number of edge excitations inside the box, which is consistent with Table. \ref{5/2Table1}. }\label{spectrum_v52mix_pbg}
\end{figure}
\begin{table}[h]
\begin{tabular}{|c||c|c|c|c|}
\hline \hline
lowest state  & $N=6$      & $N=8$       & $N=10$      & $N=12$ \\ \hline
$1$st         & $0.9147$   & $0.8147$    & $0.5458$    & $0.5108$  \\ \hline
$2$nd         & $0.0938$   & $0.0003$    & $0.0000$    & $0.0000$  \\ \hline
$3$rd         & $0.0416$   & $0.0034$    & $0.0001$    & $0.0001$  \\ \hline
$4$th         & $0.0001$   & $0.0014$    & $0.0003$    & $0.0001$  \\ \hline
$5$th         & $0.0140$   & $0.0007$    & $0.0001$    & $0.0000$  \\
\hline \hline
\end{tabular}
\caption{
Wave functions' overlaps between the destabilizing state $\psi_{d}$ with pure Coulomb interaction, and the lowest five states in the same subspace $\Delta M_{2}$ with $\lambda = 0.5$ mixed Hamiltonian. $\Delta M_{2} \equiv M_{2} - M_{P}$ is the angular momentum number's change compared to Pfaffian state's angular momentum $M_{P}=N(2N-3)/2$ for the states which initially destabilize the Pfaffian state at the smoother confinement side. The strength of confining potential is $d/l_{B} = 0.7$. The destabilizing state wave function $\psi_{d}$ has the largest overlap with the lowest state with $\lambda = 0.5$ mixed Hamiltonian for the $N = 6$, $8$, $10$ and $12$ systems.}
\label{5/2Table2}
\end{table}
If we tune $\lambda$ from $1$ (pure three-body Hamiltonian) to $0.5$,  the edge excitations can still be distinguished from the spectra, because they are separated from the bulk ones as shown in Fig. \ref{spectrum_v52mix_pbg}.

\begin{figure}[h]
\includegraphics[width=8cm]{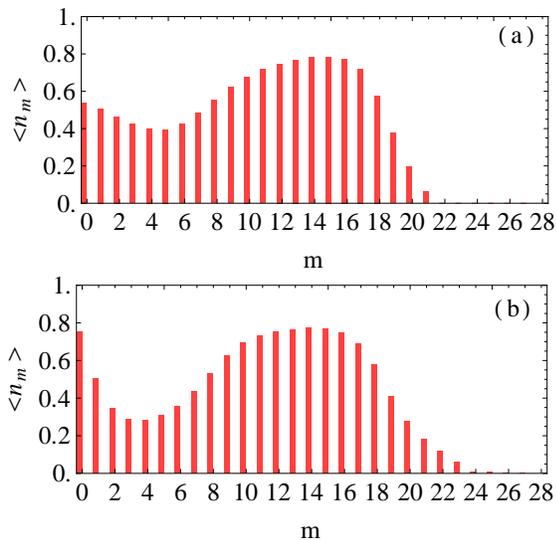}
\caption{(Color online) Orbital occupation numbers $\left\langle n_{m}\right\rangle $ of (a) the Pfaffian state and (b) the initial destabilizing state at $\Delta M = 6$ for $12$-electron system when $d/l_{B} = 0.7$. The destabilizing state occupies $2$ more orbitals than the Pfaffian state. In exact diagonalization calculation's results, the occupation number of any orbital is nonzero. But when we count how many orbitals a state occupies, we neglect the orbitals with very small occupation number ($\left\langle n_{m}\right\rangle < 0.01$). So we say there are $22$ orbitals occupied in (a) and $24$ orbitals occupied in (b).}\label{mdis52}
\end{figure}

Then we calculate wave functions' overlaps between the destabilizing state $\psi_{d}$ at the smoother confinement side (with pure Coulomb Hamiltonian), and the states with the $\lambda = 0.5$ mixed Hamiltonian at the same subspace $\Delta M$ ($d / l_{B} = 0.7$). As shown in Table \ref{5/2Table2}, the destabilizing state $\psi_{d}$ has much larger overlap with the lowest state than any other states with mixed Hamiltonian. This indicates that as $\lambda$ is tuned from $0$ to $0.5$, the destabilizing state adiabatically evolves to the lowest state with the $\lambda = 0.5$ mixed Hamiltonian, and there is no energy level crossing in this process. 
From this we conclude that this destabilizing state is an edge state, and the instability at the smoother side of the confining potential is an edge instability. 

Earlier work\cite{wan06} found that the spectrum of the (neutral) EMF mode is very close to being linear, while the (charged) bosonic EMP mode deviates from linear dispersion and bends {\em downward} as momentum increases; we thus expect this destabilizing state to be an EMP mode. This expectation is further supported by the following observation.
For systems with particle number from $6$ to $12$ (even numbers), the lowest state in subspace $\Delta M$ occupies $1$ or $2$ more orbitals than the Pfaffian state (as shown in Fig. \ref{mdis52}). The EMF wave functions constructed in Ref. \onlinecite{edgeBasis} tell us in subspaces $\Delta M > 1$ EMF excitations at least occupy $3$ more orbitals than the Pfaffian state; while the EMP excitations allow the electrons to occupy less orbitals. Therefore, the destabilizing state with pure Coulomb Hamiltonian which adiabatically evolves into the lowest state with mixed Hamiltonian, is also an EMP of the pure Coulomb Hamiltonian. By the calculation and argument above, we conclude that the state destabilizing Pfaffian state at the smoother confinement side is a Pfaffian state's EMP excitation (instead of ESW as the Abelian FQH states or EMF). Since the critical parameter $d_{c2}/l_{B}$ of this edge reconstruction has no big change as the particle number increases from $6$ to $12$ as shown in Table \ref{5/2Table}, we predict that $d_{c2}/l_{B}$ is between $0.5$ and $1$ in the thermodynamic limit. The finite Zeeman coupling will further support polarized edge excitations to reconstruct the Pfaffian state with changing confinement.

\section{CONCLUDING REMARKS}
In this paper, we have studied the low-energy excitations and edge reconstructions of FQH liquids with spin degrees of freedom. For the Abelian FQH liquids at the filling factors $\nu = 1/3$, $2/5$ and $2/3$, we find that spin plays a prominent role in edge instabilities, at least when the Zeeman splitting of $g$ factor is tuned to be sufficiently small. These results are also relevant to systems with other internal degrees of freedom (often referred to as pseudospins), including systems with valley degeneracy like graphene or Si, and multi-layered systems.

On the other hand, for non-Abelian Moore-Read Pfaffian state that may describe the FQH state at $\nu = 5/2$, we find that spin plays no role in its instability in finite size systems. Our results thus suggest that if the Moore-Read Pfaffian state is realized at $\nu = 5/2$, it is likely to be spin-polarized not only in the bulk,\cite{feiguin09} but also at the edge. Furthermore we clarified that its instability is triggered by softening of edge magnetoplasmon excitation on the smoother side of the confining potential, while on the sharp side it results from competition with other bulk states, in agreement with earlier study.\cite{wan08} We need to caution though within the way we model the 5/2 FQH liquid, we have not been able to access the anti-Pfaffian state, which has a more complicated edge structure. We cannot say anything about the role spin plays there, should that state be the one actually realized experimentally.

We conclude by stating that building upon earlier theoretical and numerical works, we have shown spins (and possibly other internal degrees of freedom) play an active role at the edge of many FQH liquids, including triggering their instabilities. They deserve more theoretical and especially experimental studies in the future.

\section*{Acknowledgements}
This work was supported by DOE Grant No. DE-SC0002140, Z. X. H. is also supported by NSFC No. 11274403 and start-up fund in ChongQing University.


\begin{thebibliography}{50}
\bibitem{2/3noteI}
The reason is that $2/3$ Halperin state is the zero energy state of a hard core interaction with inter-layer $V_{m=0,1} > 0$ and all the other pseudo-potential $V_{m} = 0$, which destroys $SU$($2$) symmetry. By further calculation, we find the $2/3$ Halperin wave function is very sensitive to the change of the hard core interaction. If we use the hard core interaction with inter-layer $V_{0} = 1$, $V_{1} = 0.5$ (close to the same ratio of Coulomb interaction), and just increase intra-layer $V_{1}$ from $0$ to $0.1$, the new ground states will only have overlaps $0.5758$, $0.0954$, $0.0033$ (for $6$-, $8$- and $10$-electron systems respectively) with the Halperin state.
\end{thebibliography}
\end{document}